\documentclass[journal]{IEEEtran}
\usepackage[T1]{fontenc}

\usepackage{graphicx}
\usepackage{amssymb,amsmath,amsfonts,amsthm}
\usepackage{mathrsfs}
\usepackage{cite}
\usepackage{array}
\usepackage{tabularx}
\usepackage[table]{xcolor}
\usepackage{color}
\usepackage{mathtools}
\usepackage{graphicx}
\usepackage{caption}
\usepackage{subcaption}
\usepackage{mathtools}
\usepackage{tikz,pgf}
\usetikzlibrary{calc,positioning,mindmap,trees,decorations.pathreplacing}
\usepackage{bbm}
\usepackage[utf8]{inputenc}
\usepackage{algorithm}

\IEEEoverridecommandlockouts
\ifCLASSINFOpdf
\else
\fi
\usepackage[cmintegrals]{newtxmath}
\begin{document}
	\title{SPSA-Based Successive Beamforming for Mobile Satellite Receivers with Phased Arrays}
	
	\author{Zheng~Chen,~\IEEEmembership{Member,~IEEE},~and~Håkan Johansson,~\IEEEmembership{Senior Member,~IEEE}
		\thanks{Z.~Chen and H. Johansson are with the Department of Electrical Engineering (ISY), Link\"{o}ping University,  58183 Link\"{o}ping, Sweden. Email: \{zheng.chen, hakan.johansson\}@liu.se.}
		\thanks{This work was supported by VINNOVA (Swedish Agency for Innovation Systems).}}
	
	\maketitle

\begin{abstract}	
Efficient and low-complexity beamforming design is an important element of satellite communication systems with mobile receivers equipped with phased arrays. In this work, we apply the simultaneous perturbation stochastic approximation (SPSA) method with successive sub-array selection for finding the optimal antenna weights that maximize the received signal power at a uniform plane array (UPA). The proposed algorithms are based on iterative gradient approximation by injecting some carefully designed perturbations on the parameters to be estimated. Additionally, the successive sub-array selection technique enhances the performance of SPSA-based algorithms and makes them less sensitive to the initial beam direction. 
Simulation results show that our proposed algorithms can achieve efficient and reliable performance even when the initial beam direction is not well aligned with the satellite direction.
\end{abstract}

\section{Introduction}
Satellite communication (SatCom) has been a long-standing research field for several decades. Recently, it has regained the spotlight because of the intensive discussions on integrating SatCom infrastructure into mobile networks to achieve seamless and flexible wireless connectivity in 5G and beyond \cite{del2002next}. The related application scenarios are particularly meaningful for providing mobile coverage in those areas that are under-served by terrestrial cellular base stations, such as oceans and deserted areas.

Phased array antennas are commonly considered for SatCom on-the-move applications where mobile terminals are communicating with satellites. A phased array can steer in a certain direction by electronically adjusting the phase shifts between the antenna elements. When a phased array on a moving platform receives satellite signals, the beam pointing direction needs to match the direction of arrival of the satellite signals to maximize the received signal power. This procedure is referred to as beam tracking. Though the fundamental theories of phased arrays are well investigated, how to exploit the structure of phased arrays to achieve efficient beam tracking with low-cost and low-complexity algorithms for SatCom on-the-move scenario remains a meaningful topic.

Simultaneous perturbation stochastic approximation (SPSA) is a powerful and effective stochastic optimization tool for solving multivariate optimization problems without knowing exactly the gradient of the objective function with respect to the parameters being optimized \cite{spsa-mle, spsa, spsa-implementation}. This method has been applied in various engineering scenarios that involve solving an optimization problem without explicitly knowing the model, such as source seeking in mobile robot systems \cite{source-seeking}. For beam tracking in SatCom scenario with phased arrays, the parameters to be optimized are the phase shifts of the antenna elements. When the received signal power at the antenna array is the only available knowledge, SPSA can be a potentially effective solution for finding the optimal phase shifts by using gradient approximation. In \cite{beamforming-SPSA}, a receive beamforming algorithm based on SPSA method is proposed for a uniform linear array (ULA).
Similar algorithms for uniform plane array (UPA) have been proposed in \cite{fast-beamforming-mobile, fast-beamforing-array, low-cost-bf,zero-knowledge}.
In  \cite{beam-tracking-jsac}, both mechanical and electrical adjustments are considered for beam tracking in UAV-mounted SatCom, where the electrical adjustment algorithm is developed by using SPSA with the array structure information. Recently, SPSA methods have been adopted for training end-to-end communication systems with deep neural networks \cite{BP-air}.

Nevertheless, all the prior works consider the original form of the SPSA where two noisy measurements with perturbed parameters need to be simultaneously obtained in each iteration. This assumption can be unrealistic in practical scenarios where the channel conditions vary between two measurements. Another problem is that SPSA-based algorithms might fail to converge to the global optimal solution if the objective function is not convex or quasi-convex. In SatCom scenarios with phased arrays at the receiver side, the objective is to align the beam direction with the satellite direction. When the initial beam direction is far from the satellite direction, SPSA-based algorithms might converge to one of the local optimal points in the side lobes of the antenna array.

In this paper, we utilize both the original two-point form and the modified one-point form of the SPSA method and propose two angle-domain SPSA algorithms with successive sub-array selection for mobile satellite receivers. Simulation results showed that our proposed algorithms provide an effective and low-cost solution for receive beamforming in SatCom systems with phased arrays. Particularly, the successive sub-array selection design makes the SPSA-based algorithms less sensitive to the initial beam direction and achieves more robust performance under practical conditions.   
 
\section{System Model}
We consider a satellite communication system where a moving vehicle receives signals from a single-antenna geostationary satellite. The vehicle is equipped with a two-dimensional UPA, which contains $M\times N$ uniformly spaced antennas with antenna spacing equal to $d$. The UPA has only one RF chain, which corresponds to the single channel receiver scenario.
Let $\alpha$ and $\beta$ denote the direction of arrival (DoA) of the satellite signals seen at the UPA, where $\alpha$ is the azimuth angle and $\beta$ is the elevation angle.

\subsection{Channel Model}
Assuming zero-band signal with carrier frequency $f_c$, the frequency response at the $(m,n)$-th antenna element is:
\begin{equation}
A(m,n)\!=\!\exp\!\bigg(j \frac{2\pi d f_c}{c}\big[(m-1)\sin\alpha\! \cos\beta
+(n-1)\sin\alpha\!\sin\beta\big]\!\bigg),
\end{equation}
where $c$ is the speed of light, $m=1,\ldots,M$, and $n=1,\ldots,N$.

Let $\mathbf{A}(\alpha, \beta)$ represent the $M\times N$ array gain matrix when the satellite signal arrives at the UPA with the DoA $(\alpha, \beta)$.
Considering the fact that the channel between the satellite and a mobile vehicle on the ground is most likely to be line-of-sight (LoS), we consider a similar channel model as in \cite{beam-tracking-jsac}. The $M\times N$ channel matrix is be modeled as  
\begin{equation}
\mathbf{H}=C\cdot \exp(j\phi_0)\mathbf{A},
\end{equation}
where $C$ is a complex-valued constant parameter related to the signal power attenuation in the environment, and $\phi_0$ denotes the phase of the arrived signal at the first antenna element.
The vector form of the channel gain can be written as $\mathbf{h}=\text{vec}(\mathbf{H})$, where the dimension of $\mathbf{h}$ is $MN\times 1$.

We assume that the UPA contains phase shifters without amplification capability. Let  $\mathbf{w}=[w_1,\ldots, w_{MN}]^T$ denote the vector of antenna weights, we have $|w_i|=1, \forall i$. Each antenna weight corresponds to a certain phase shift, i.e., $w_i=\exp(j\phi_i)$, where $\phi_i$ is the phase shift of the $i$-th antenna element. 
The received signal at the output of the antenna array is
\begin{equation}
y=\mathbf{w}^{H}\mathbf{h} s+\mathbf{w}^{H}\mathbf{n},
\end{equation}
where $s$ denotes the transmitted data symbol from the satellite, and $\mathbf{n}$ denotes the size $MN\times 1$ noise vector added at the input of the UPA. 
The instantaneous received signal power is
	\begin{equation}
	P=|y|^2=\mathbf{w}^{H}\mathbf{R}_s \mathbf{w}+\mathbf{w}^{H}\mathbf{R}_{n}\mathbf{w},
	\end{equation}
where $\mathbf{R}_s =\mathbf{h} s s^{H}\mathbf{h}^{H}$ and $\mathbf{R}_n =\mathbf{n} \mathbf{n}^{H}$ are covariance matrices of the signal and noise. 

If we have perfect knowledge of the channel vector $\mathbf{h}$ (or the DoA), it is well known that the received signal power or the signal-to-noise ratio is maximized when a matched filter is applied. Then, the optimal antenna weight vector is $\mathbf{w}=\text{vec}(\mathbf{A}(\alpha,\beta))$, given the DoA of the satellite signals. In reality, accurate DoA estimation usually requires advanced RF chains with high computational complexity. Therefore, in this work, we propose several low-cost zero-knowledge algorithms for single channel receive beamforming without the need of advanced hardware and signal processing units. The theory behind the proposed algorithms is simultaneous perturbation stochastic approximation (SPSA) method.

\subsection{Preliminaries on SPSA}
SPSA is an effective multivariate optimization method when the gradient of the objective function with respect to the optimization parameters is not directly available\cite{spsa-implementation}.
Consider the problem of minimizing a loss function $L(\mathbf{x})$ which is a scalar measure of the system performance and $\mathbf{x}$ is a $p$-dimensional vector containing the parameters to be optimized. 
SPSA is an iterative optimization method that contains three main steps:
\begin{enumerate}
	\item In each iteration $k$, generate some random perturbation vector $\boldsymbol{\delta}_k$ in the following form
	\begin{equation}
	\boldsymbol{\delta}_k=\frac{c \Delta_k}{(k+1)^\gamma},
	\end{equation}
	where $c$ and $\gamma$ are constant parameters, and $\Delta_k$ is a random vector whose elements are independently generated from some probability distribution. 
	\item  Obtain two noisy measurements of the loss function $L(\hat{\mathbf{x}}_k+\boldsymbol{\delta}_k)$ and $L(\hat{\mathbf{x}}_k-\boldsymbol{\delta}_k)$. The gradient of the loss function is approximated by 
	\begin{equation}
	\hat{g}(\hat{\mathbf{x}}_k)=\frac{L(\hat{\mathbf{x}}_k+\boldsymbol{\delta}_k)-L(\hat{\mathbf{x}}_k-\boldsymbol{\delta}_k)}{2\boldsymbol{\delta}_k}.
	\end{equation}
	\item Update the estimated parameter by
	\begin{equation}
	\hat{\mathbf{x}}_{k+1}=\hat{\mathbf{x}}_k-a_k\hat{g}(\hat{\mathbf{x}}_k),
	\end{equation}
	where $a_k$ is the step size (sequence gain).
\end{enumerate}
When the parameters are appropriately chosen following the guidelines in \cite{spsa-implementation}, this iterative algorithm  will effectively converge to a local optimum point.

\section{SPSA-Based Successive Receive Beamforming}
\label{sec:SPSA}
Existing applications of the SPSA method for SatCom systems with phased arrays usually consider approximating the gradient of the received signal power subject to the phase shifts. For the case with a uniform linear array, an SPSA-based receive beamforming algorithm is proposed in \cite{beamforming-SPSA} by considering phase-domain perturbation.
The extension of this method to a two-dimensional UPA is not straightforward, because the phase difference between the received signals at two neighboring antennas is not always proportional to the distance between them. Another problem with the SPSA-based method is that the algorithm performance is strongly affected by the initial antenna weights. All the related works have to assume that the real DoA angles $\alpha$ and $\beta$ are very small, or the initial beam direction is very close to the satellite direction \cite{beamforming-SPSA, beam-tracking-jsac}. Otherwise, the algorithm might converge to a local optimum point within the side lobes instead of the main lobe in the radiation pattern of the antenna array.

To resolve this problem, we propose a successive beamforming algorithm using the angle-domain SPSA method and consider the DoA angles as the optimization parameters. The key design point is to choose a subset of antennas to apply the SPSA-based beamforming algorithm, and gradually increase the number of antennas being utilized to improve the accuracy of the beam pointing direction. 
Compared to the existing designs, our proposed design is much less sensitive to the initial antenna weights, which makes it more robust under realistic conditions.

\subsection{Angle Domain Successive SPSA}
\label{sec:SPSA1}
Let $\boldsymbol{\Phi}=[\alpha, \beta]$ be the parameter vector to be estimated, and $\boldsymbol{\delta}=[\delta_1, \delta_{2}]$ be the perturbation vector in each iteration. 
We define the following function of the DoA angles and the antenna index $(m,n)$:
\begin{equation}
	\begin{split}
	f(\alpha,\beta, m,n)=\exp\bigg(j 2\pi d\frac{f_c}{c}\big[(m-1)\sin\alpha \cos\beta
\\+(n-1)\sin\alpha \sin\beta\big]\bigg).
	\end{split}	
	\label{eq:function_alpha_beta}
\end{equation}
The successive SPSA algorithm works as follows.
\begin{enumerate}
	\item \textbf{Initialization.} Set the initial estimated angle vector as $\hat{\boldsymbol{\Phi}}_0=[\alpha_0,\beta_0]$. 
	
	\item \textbf{Antenna sub-array selection.} For the $l$-th global iteration, choose a sub-array of $M_l\times N_l$ antennas, where $1<M_l\leq M$, $1<N_l\leq N$, $M_l>M_{l-1}$ and $N_l>N_{l-1}$. The corresponding antenna weights are obtained by $\mathbf{w}(m,n)=f(\alpha_0,\beta_0, m,n)$, $\forall m=\{1,\ldots, M_l\}, n=\{1,\ldots, N_l\}$. Then, within the $l$-th global iteration, run $K$ local iterations of stochastic approximation with perturbed parameters.

	\item \textbf{Stochastic Approximation.} For each local iteration $k=1,\ldots, K$, set the perturbation vector as
	\begin{equation} 
	\boldsymbol{\delta}_k=\frac{b }{(k+1)^\Omega}\cdot [x_1,x_2],
	\label{eq:pertub_vector}
	\end{equation}
	where $x_1, x_2\in\{+1,-1\}$ are binary variables with equal probability $1/2$, $b$ and $\Omega$ are constant parameters chosen by following the guidelines in \cite{spsa-implementation}. 
	The estimated gradient of the received signal power is
	\begin{equation}
	\hat{g}({\hat{\boldsymbol{\Phi}}_k})=\frac{P(\hat{\boldsymbol{\Phi}}_k+\boldsymbol{\delta}_k)-P(\hat{\boldsymbol{\Phi}}_k-\boldsymbol{\delta}_k)}{2\boldsymbol{\delta}_k}.
	\label{eq:estigrad}
	\end{equation}	
	The received signal power with the perturbed angle vector is measured by setting the antenna weight at the $(m,n)$-th element as $\mathbf{w}(m,n)=f(\tilde{\alpha},\tilde{\beta}, m,n)$.
	where $\tilde{\alpha}$ and $\tilde{\beta}$ are the perturbed DoA angles, corresponding to the first and second elements in $\hat{\boldsymbol{\Phi}}_k\pm\boldsymbol{\delta}_k$, respectively.
	\item Adjust the estimated angle vector in the next iteration:
	\begin{equation}
	\hat{\boldsymbol{\Phi}}_{k+1}=\hat{\boldsymbol{\Phi}}_k+\eta_k\hat{g}({\hat{\boldsymbol{\Phi}}_k}),
	\label{eq:update_phi}
	\end{equation}
	where $\eta_k=a/(\zeta+k)^{\xi}$ is the step size, with $a$ and $\xi$ being priorly chosen parameters.\footnote{Note that the plus sign in \eqref{eq:update_phi} is because the objective is to maximize the received signal power instead of minimize it.}
	\item Update the antenna weight at the $(m,n)$-th element by $\mathbf{w}(m,n)=f(\hat{\alpha},\hat{\beta}, m,n)$,
	where $\hat{\alpha}=\hat{\boldsymbol{\Phi}}_{k+1}(1)$ and $\hat{\beta}=\hat{\boldsymbol{\Phi}}_{k+1}(2)$ are the estimated DoA angles for the next local iteration.
	\item After executing steps $(3-5)$ $K$ times, the estimated DoA $\hat{\alpha},\hat{\beta}$ in the last local iteration will be returned to step $2$, as the initial DoA angles for the next global iteration with increased size of the sub-array.
	\item Repeat this procedure until the termination criterion is satisfied, e.g., $|P(\hat{\boldsymbol{\Phi}}_{k+1})-P(\hat{\boldsymbol{\Phi}}_k)|<\epsilon$, where $\epsilon$ is a pre-defined threshold.
\end{enumerate}

\subsection{Modified One-Point SPSA}
\label{sec:SPSA3}
Though SPSA is a very powerful method for derivative-free multivariate optimization problems, in a practical communication system, two different measurements can never been done simultaneously in the strict sense. Only when the environment remains unchanged while two measurements are obtained sequentially, these two measurements can be approximately considered as simultaneous. 

A one-point form of the SPSA method is proposed in \cite{spsa-one}, where only one noisy measurement is required in each iteration. Recall that for the perturbation vector defined in \eqref{eq:pertub_vector}, $x_1, x_2\in\{+1,-1\}$ are binary variables with equal probabilities. The perturbation vector has four discrete directions, which are $[+1, +1], [+1, -1], [-1, +1]$, and $[-1, -1]$. Due to the random perturbation direction, it is possible that the perturbation is set toward undesired direction consecutively in many iterations, which results in longer convergence time. 
We propose a modified one-point SPSA algorithm which combines the advantages of the two-point and one-point SPSA methods. The new algorithm takes only one noisy measurement in each iteration, while the perturbation direction is set to be cyclic. Based on the original successive SPSA algorithm described in Section~\ref{sec:SPSA1}, the main differences are as follows.
\begin{itemize}
	\item At each local iteration $k=1,\ldots,K$, we set the perturbation vector by \eqref{eq:pertub_vector}, where $[x_1,x_2]$ periodically takes values from $\{[+1, +1], [+1, -1], [-1, +1], [-1, -1]\}$ in a cyclic way. The antenna weights are adjusted by $\mathbf{w}(m,n)=f(\tilde{\alpha},\tilde{\beta}, m,n)$ with
	\begin{equation}
	\begin{cases}
	&\tilde{\alpha}=\hat{\boldsymbol{\Phi}}_k(1)+\boldsymbol{\delta}_k(1)  \\
	&\tilde{\beta}=\hat{\boldsymbol{\Phi}}_k(2)+\boldsymbol{\delta}_k(2) 
	\end{cases}
	\end{equation} 
	One measurement of the received signal power $\tilde{P}_k$ is obtained with the new antenna weights.
	\item 
	The estimated gradient of the received signal power is
	\begin{equation}
	\hat{g}({\hat{\boldsymbol{\Phi}}_k})=\frac{\tilde{P}_k-\tilde{P}_{k-1}}{\boldsymbol{\delta}_k-\boldsymbol{\delta}_{k-1}},
	\end{equation}	
	where $\tilde{P}_k=P(\tilde{\Phi}_k)$ with $\tilde{\Phi}_k=[\tilde{\alpha},\tilde{\beta}]$.
	If the value of $x_1$ in iteration $k$ is the same as in iteration $k-1$, then the first element of $\hat{g}({\hat{\boldsymbol{\Phi}}_k})$ is set to be zero to avoid an infinite-valued element in $\hat{g}({\hat{\boldsymbol{\Phi}}_k})$. The same rule applies for the value of $x_2$ and the second element of $\hat{g}({\hat{\boldsymbol{\Phi}}_k})$.
\end{itemize}

\section{Simulation Results}
In this section, we present and compare the performance of the proposed algorithms by performing Matlab-based simulations. We consider an UPA with $8\times 8$ antennas, with uniform antenna spacing $d=\lambda_c/2=0.75$ cm, where $\lambda_c=c/f_c$ is the wavelength at carrier frequency $f_c=20$ GHz. 
The parameters in the SPSA algorithm are chosen as $a=8\times 10^{-6}$, $b=\pi/18$, $\Omega=0.101$, $\zeta=0.602$, which follow the guidelines in \cite{spsa-implementation}. The signal-to-noise ratio at the receiver side is $20$\,dB.  In the initialization step, the estimated DoA angles are $\alpha_0=0^\circ$ and $\beta_0=0^\circ$. 
We consider the normalized received signal power (NRSP) as the performance metric, which is the ratio between the received signal power and the maximum power that can be achieved in the optimal case.

For the ease of illustration, we name the proposed algorithms as follows.
\begin{itemize}
	\item \textit{SPSA-two}: the original two-point SPSA algorithm described in Section \ref{sec:SPSA1}.
	\item \textit{SPSA-one}: the modified one-point SPSA algorithm described in Section \ref{sec:SPSA3}
\end{itemize}

\begin{figure}
     \centering
	\begin{subfigure}[b]{0.5\textwidth}
		\centering
		\includegraphics[width=\textwidth]{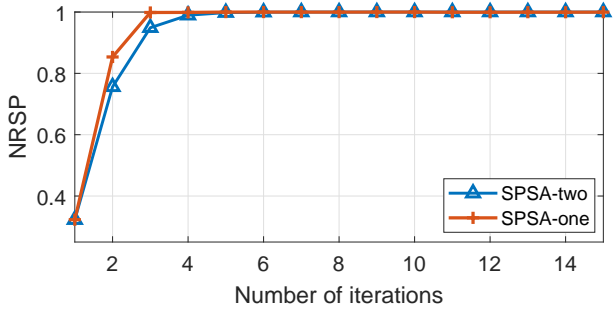}
		\caption{The DoA angles are $\alpha=8^\circ$ and $\beta=6^\circ$. }
		\label{fig:case1}
	\end{subfigure}\hfill
	\begin{subfigure}[b]{0.5\textwidth}
		\centering
		\includegraphics[width=\textwidth]{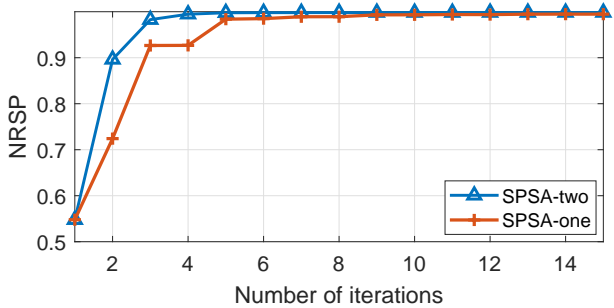}
		\caption{The DoA angles are $\alpha=6^\circ$ and $\beta=8^\circ$. }
		\label{fig:case2}
	\end{subfigure}
\caption{Performance comparison between the two-point and one-point SPSA algorithms without successive sub-array selection. }
\label{fig:spsa-two-one-comparison}
\end{figure}

In Fig.~\ref{fig:spsa-two-one-comparison}, we show the NRSP performance of the two-point and one-point SPSA algorithms without successive sub-array selection. Depending on the specific DoA angles, one-point SPSA might achieve slightly better or worse performance than the two-point algorithm. However, the main advantage of the one-point SPSA method lies in the fact that only one measurement instead of two is required in each iteration. Considering this effect, we conclude that the modified one-point SPSA algorithm achieves comparable performance as the two-point version with only half of the measurements.

\begin{figure}[t!]
	\centering
	\includegraphics[width=\columnwidth]{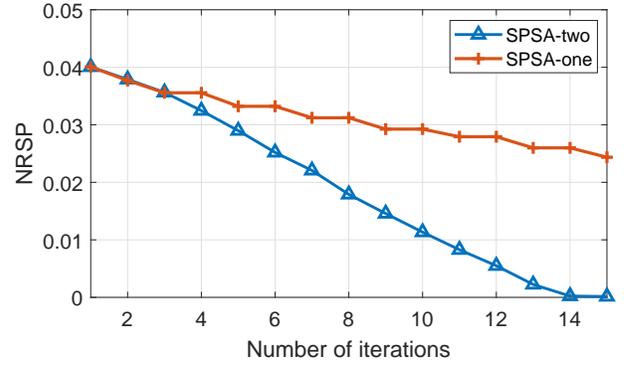}
	\caption{NSRP of the two-point and one-point SPSA algorithms without successive sub-array selection.  The real DoA angles are $\alpha=20^\circ$ and $\beta=15^\circ$. }
	\label{fig:spsa-bad-case}
\end{figure}  
As mentioned at the beginning of Section~\ref{sec:SPSA}, the performance of SPSA-based receive beamforming algorithms can be strongly affected by the initial  DoA angles $\alpha_0$ and $\beta_0$. 
In Fig. \ref{fig:spsa-bad-case}, we show one example where the original SPSA algorithms (without successive sub-array selection) fail to converge to the optimal point. This happens when the real DoA angles are too far from the initial beam direction. In this case, SPSA-based algorithms can only converge to the nearest local optimum in the side lobes instead of the main lobe.

\begin{figure}[t!]
	\centering
	\includegraphics[width=\columnwidth]{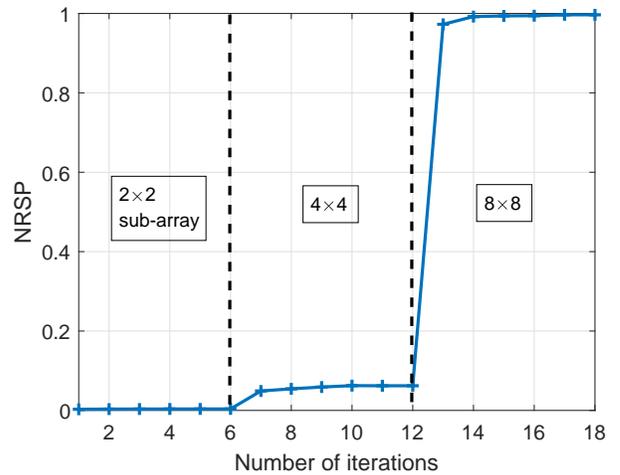}
	\caption{Performance of the one-point SPSA algorithm with successive sub-array selection. The real DoA angles are $\alpha=20^\circ$ and $\beta=15^\circ$. }
	\label{fig:successive}
\end{figure}  
In Fig.~\ref{fig:successive}, we show the NRSP of the successive SPSA-based receive beamforming algorithm. For the first six iterations, a sub-array of dimension $2\times2$ is selected. Then the selected sub-array increases to $4\times 4$ for the next six iterations. At last, the full antenna array is utilized. The NRSP is obtained as the receive signal power divided by the maximum received signal power with $8\times 8$ full antenna array. For this reason, the NRSP values obtained with $2\times 2$ and $4\times 4$ sub-arrays are much smaller than the case with $8\times 8$ array.
We see that with the same DoA angles as in Fig.~\ref{fig:spsa-bad-case}, our proposed successive beamforming algorithm can still converge to the global optimal point. This result shows that the successive sub-array selection technique can improve the robustness and effectiveness of the SPSA-based receive beamforming algorithms.

\section{Conclusions}
In this work, we proposed two receive beamforming algorithms using the SPSA method and successive antenna sub-array selection for mobile satellite receivers equipped with a two-dimensional phased array. The proposed algorithms were shown to be robust and efficient at finding the optimal beamforming coefficients, especially when the DoA angles of the satellite signals are relatively large. 


\begin{thebibliography}{10}
	\providecommand{\url}[1]{#1}
	\csname url@samestyle\endcsname
	\providecommand{\newblock}{\relax}
	\providecommand{\bibinfo}[2]{#2}
	\providecommand{\BIBentrySTDinterwordspacing}{\spaceskip=0pt\relax}
	\providecommand{\BIBentryALTinterwordstretchfactor}{4}
	\providecommand{\BIBentryALTinterwordspacing}{\spaceskip=\fontdimen2\font plus
		\BIBentryALTinterwordstretchfactor\fontdimen3\font minus
		\fontdimen4\font\relax}
	\providecommand{\BIBforeignlanguage}[2]{{%
			\expandafter\ifx\csname l@#1\endcsname\relax
			\typeout{** WARNING: IEEEtran.bst: No hyphenation pattern has been}%
			\typeout{** loaded for the language `#1'. Using the pattern for}%
			\typeout{** the default language instead.}%
			\else
			\language=\csname l@#1\endcsname
			\fi
			#2}}
	\providecommand{\BIBdecl}{\relax}
	\BIBdecl
	
	\bibitem{del2002next}
	E.~Del~Re and L.~Pierucci, ``Next-generation mobile satellite networks,''
	\emph{IEEE Communications Magazine}, vol.~40, no.~9, pp. 150--159, 2002.
	
	\bibitem{spsa-mle}
	J.~C. Spall, ``A stochastic approximation technique for generating maximum
	likelihood parameter estimates,'' in \emph{1987 American Control
		Conference}.\hskip 1em plus 0.5em minus 0.4em\relax IEEE, 1987, pp.
	1161--1167.
	
	\bibitem{spsa}
	J.~C. {Spall}, ``Multivariate stochastic approximation using a simultaneous
	perturbation gradient approximation,'' \emph{IEEE Transactions on Automatic
		Control}, vol.~37, no.~3, pp. 332--341, 1992.
	
	\bibitem{spsa-implementation}
	J.~C. Spall, ``Implementation of the simultaneous perturbation algorithm for
	stochastic optimization,'' \emph{IEEE Transactions on aerospace and
		electronic systems}, vol.~34, no.~3, pp. 817--823, 1998.
	
	\bibitem{source-seeking}
	E.~{Ramírez-Llanos} and S.~{Martínez}, ``Stochastic source seeking for mobile
	robots in obstacle environments via the {SPSA} method,'' \emph{IEEE
		Transactions on Automatic Control}, vol.~64, no.~4, pp. 1732--1739, 2019.
	
	\bibitem{beamforming-SPSA}
	F.~G. {Zhang}, W.~M. {Jia}, W.~{Jin}, and M.~L. {Yao}, ``Beamforming algorithm
	based on {SPSA} for mobile satellite receiver,'' \emph{Electronics Letters},
	vol.~48, no.~22, pp. 1379 --1380, October 2012.
	
	\bibitem{fast-beamforming-mobile}
	M.~{Fakharzadeh}, S.~H. {Jamali}, P.~{Mousavi}, and S.~{Safavi-Naeini}, ``Fast
	beamforming for mobile satellite receiver phased arrays: Theory and
	experiment,'' \emph{IEEE Transactions on Antennas and Propagation}, vol.~57,
	no.~6, pp. 1645--1654, June 2009.
	
	\bibitem{fast-beamforing-array}
	{Chen Sun}, A.~{Hirata}, T.~{Ohira}, and N.~C. {Karmakar}, ``Fast beamforming
	of electronically steerable parasitic array radiator antennas: theory and
	experiment,'' \emph{IEEE Transactions on Antennas and Propagation}, vol.~52,
	no.~7, pp. 1819--1832, July 2004.
	
	\bibitem{low-cost-bf}
	P.~{Mousavi}, M.~{Fakharzadeh}, S.~H. {Jamali}, K.~{Narimani}, M.~{Hossu},
	H.~{Bolandhemmat}, G.~{Rafi}, and S.~{Safavi-Naeini}, ``A low-cost ultra low
	profile phased array system for mobile satellite reception using
	zero-knowledge beamforming algorithm,'' \emph{IEEE Transactions on Antennas
		and Propagation}, vol.~56, no.~12, pp. 3667--3679, 2008.
	
	\bibitem{zero-knowledge}
	M.~{Hossu}, S.~H. {Jamali}, P.~{Mousavi}, K.~{Narimani}, M.~{Fakharzadeh}, and
	S.~{Safavi-Naeini}, ``Zero-knowledge adaptive beamforming using analog signal
	processor for satellite tracking applications with an experimental comparison
	to a digital implementation,'' \emph{IEEE Transactions on Aerospace and
		Electronic Systems}, vol.~46, no.~3, pp. 1533--1543, 2010.
	
	\bibitem{beam-tracking-jsac}
	J.~{Zhao}, F.~{Gao}, Q.~{Wu}, S.~{Jin}, Y.~{Wu}, and W.~{Jia}, ``Beam tracking
	for {UAV} mounted {SatCom} on-the-move with massive antenna array,''
	\emph{IEEE Journal on Selected Areas in Communications}, vol.~36, no.~2, pp.
	363--375, 2018.
	
	\bibitem{BP-air}
	V.~{Raj} and S.~{Kalyani}, ``Backpropagating through the air: Deep learning at
	physical layer without channel models,'' \emph{IEEE Communications Letters},
	vol.~22, no.~11, pp. 2278--2281, 2018.
	
	\bibitem{spsa-one}
	J.~C. Spall, ``A one-measurement form of simultaneous perturbation stochastic
	approximation,'' \emph{Automatica}, vol.~33, no.~1, pp. 109--112, 1997.
	
\end{thebibliography}

\end{document}